\documentclass[preprint,12pt]{elsarticle}

\usepackage{amssymb}
\usepackage{amsmath, amsfonts}
\usepackage{graphicx}
\usepackage{color}
\usepackage[usenames,dvipsnames,svgnames]{xcolor}
\usepackage[colorlinks=true,
      linkcolor=red,
      urlcolor=gray,
      citecolor=blue]{hyperref}
 \usepackage{hyperref}

\journal{New Astronomy}

\begin{document}

\begin{frontmatter}



\title{Matter power spectrum in a power-law $f(G)$ gravity}


\author{Albert Munyeshyaka} 

\affiliation{organization={Rwanda Astrophysics Space and Climate Science Research Group, University of Rwanda, College of Science and Technology},
            city={Kigali},
            country={Rwanda}}
						

\author{Praveen Kumar Dhankar} 

\affiliation{organization={Symbiosis Institute of Technology, Nagpur Campus, Symbiosis International (Deemed University)},
            addressline={Pune-440008}, 
            city={Maharashtra},
            country={India}}
\author{Joseph Ntahompagaze} 

\affiliation{organization={University of Rwanda, College of Science and Technology},
            city={Kigali},
            country={Rwanda}}

\begin{abstract}
Cosmological models based on $f(G)$ gravity are efficient in fitting different observational datasets at both background and perturbation levels. This motivates the current study to take into account dynamical system analysis to investigate the matter power spectrum within the framework of modified Gauss-Bonnet gravity. After defining the dimensionless dynamical system variables for a power-law $f(G)$ model, We derive  the full system of equations governing the energy density perturbations for both matter and Gauss-Bonnet fluids using the $1+3$ covariant formalism. After solving the energy density perturbation equations, we compute the matter power spectrum. The importance of studying first order perturbations for the defined $f(G)$ model  and the relevance of different initial conditions in computing the matter power spectrum are also stressed. It is reported that matter power spectrum for $f(G)$ gravity, for a particular functional form of $f(G)$ model considered is not scale invariant as the case for General Relativity.
\end{abstract}



\begin{keyword}
Cosmology; cosmic acceleration; dark energy; dynamical system; $1+3$ formalism; Gauss-Bonnet gravity.


\end{keyword}

\end{frontmatter}



\section{Introduction}\label{intro}
In the past decades, advances in observational cosmology helped for the accurate description of the universe using the $\Lambda$CDM model. This phenomenological model appears to fit available observations such as Supernovae $I_{a}$\cite{perlmutter1999astrophys,dodelson2000dark}, Cosmic microwave background radiations anisotropy \cite{cornish2004constraining,spergel2007three}, large scale structure formation \cite{tegmark2004cosmological} to name but a few. Unfortunately the $\Lambda$CDM model based on a dynamical scalar field and introducing a positive cosmological constant, fails to describe for example the current cosmic accelerated expansion of the universe.\\ This has led to the efforts to seek for alternative theory based on the modification of standard Einstein-Hilbert action. This implies adding a functional form of  Ricci scalar or its derivatives to the  gravitaional action. 
The leading modified gravity theories includes the $f(R)$ gravity\cite{singh2020cosmological, de2010theories, sotiriou2010f}, where $R$ is the Ricci scalar. Among other modified theories of gravity includes the $f(G)$ gravity, $G$ is Gauss-Bonnet invariant. The Gauss-Bonnet invariant defined as $G\equiv R^{2}-4R^{ij}_{ij}+R^{ijkl}_{ijkl}$, where $R^{ij}_{ij}$ is the Ricci tensor and $R^{ijkl}_{ijkl}$ is the Riemann tensor, may arise naturally in the gauge theories like the Chern-Simons \cite{gomez2011standard}, Lovelock \cite{bajardi2021exact} or Born-Infeld gravity \cite{tseytlin1997non}. The cosmological properties of modified Gauss-Bonnet gravity model, in principle, explains the primordial inflationary phase and the late time accelerated expansion of the universe.\\
 The important feature of these theories is that  the field equations can be rewritten in a way making it easy to compare with GR or $\Lambda$CDM. This combination of non-linear Ricci curvature and its derivatives leads to a curvature fluid driven period of late time acceleration. \\ This so called $f(G)$ models have been studied in different context \cite{garcia2011energy,li2007cosmology,de2020tracing,benetti2018observational}.\\ The work done in \cite{de2012stability} analysed the stability of the cosmological solutions in $f(R,G)$ gravity and helped to constrain the form of the gravitational action and understanding of the behavior of the perturbations in $f(R,G)$ class of higher order theories of gravity. This can help in more precise analysis of the full spectrum of cosmological perturbations. \\ The work conducted by Munyeshyaka et.al \cite{munyeshyaka2024covariant} analysed the covariant perturbations with scalar field in modified Gauss-Bonnet gravity and found that the energy over-density decay with redshift.\\ \\ 
Different studies on the expansion history of a $f(G)$ gravity and modified theories of gravity have been conducted using different strategies for numerically solving  cosmological equations.  These studies emphasized how the results behave due to the initial conditions, the presence of singularities and oscillations in the decerelation and snap parameters \cite{bamba2012cosmic,bamba2011phantom}.\\ These concerns can be solved by using the dynamical system approach \cite{wainwright1997dynamical,bohmer2017dynamical} which  provides a relatively simple method for obtaining exact solutions and can describe the dynamics of the universe for a given $f(G)$ model. \\ The work conducted by Ashutosh \cite{singh2025dynamical} constructed a dynamical system of modified Gauss-Bonnet gravity and discussed its cosmological implications.  Ntahompagaze et. al \cite{ntahompagaze2022large} presented large scale structure power spectrum from scalar-tensor gravity using dynamical system analysis.\\ The work done in \cite{abebe2013large} studied the predicted power spectra in the context of $f(R)$ theories of gravity using both dynamical system approach for the background and solving for the matter perturbations without using the quasi-static approximation and compared the theoretical results with several SDSS data.\\ \\
 Knowing the background expansion history of the universe which are consistent with $\Lambda$CDM model helps to investigate the growth of structure. In the present work, we study the predicted power spectra in the context of $f(R,G)$ theories of gravity using both dynamical system approach for the background and solving for the matter perturbations without using the quasi-static approximation and compare the theoretical results with General relativity prediction.\\ After obtaining the perturbation equations, and using the dimensionless autonomous dynamical variables, we solve the equations to get the energy density $\Delta(z)$ which then used to compute matter power spectra for a power-law $f(G)$ model. \\\\The rest of this paper is organised as follows: In Section (\ref{sec2}), we present the mathematical framework, where cosmological equations in the context of $f(G)$ gravity and the dynamical system approach to the matter-Gauss-Bonnet fluids mixture are discussed. In Section (\ref{sec3}), we present the covariant density perturbation equations whereas in Section (\ref{sec4}) we compute the power spectrum resulting from the energy density perturbation equations and  Section (\ref{sec5}) discuss the results and concludes.
 
\section{Mathematical framework}\label{sec2}

\subsection{Cosmological equations in the context of $f(G)$ gravity}
The gravitational action involving normal matter assisted by $f(G)$ gravity is  presented as \cite{li2007cosmology,venikoudis2022late}
\begin{eqnarray}
 S=\int d^{4}x\sqrt{-g}\Big(\frac{R}{2\kappa^{2}}+\frac{f(G)}{2}+\mathcal{L}_{m}\Big)\;
 \label{eq1},
\end{eqnarray}
where $f(G)$ represents an arbitrary function depending on the Gauss-Bonnet invariant $G$, $\mathcal{L}_{m}$ is the usual matter Lagrangian and $g$ is the determinant of the metric $g^{ij}$ and $\kappa$ is the gravitational constant. 
For the case $f(G)=G$, $\int d^{4}x\sqrt{-g}G=0$, then we recover the  gravitational action for GR, $S=\int d^{4}x\sqrt{-g}\Big(\frac{R}{2\kappa^{2}}+\mathcal{L}_{m}\Big)$. In this context, eq. (\ref{eq1}) produces field equations represented as
\begin{eqnarray}
 G_{ij} \equiv R_{ij}-\frac{1}{2}g_{ij}R=8\pi G_{N}T^{tot}_{ij},\label{eq2}
\end{eqnarray}
where $T^{tot}_{ij}$ represents the energy momentum tensor of the total fluids.  For a perfect fluid, the energy momentum tensor is given by
\begin{eqnarray}
 T_{ij}=(\rho+p)u^{i}u^{j}+pg^{ij},
\end{eqnarray}
where, $\rho$ and $p$ are the energy density and isotropic pressure respectively and $u^{i}$ is the $4$-velocity. The Friedman-Robertson-Walker (FRW) metric is given by 
\begin{eqnarray}
 ds^{2}=-dt^{2}+a^{2}(t)\Big(dx^{2}+dy^{2}+dz^{2}\Big),
 \label{eq4}
\end{eqnarray}
where $a(t)$ is the scale factor governing the expansion of the Universe and it is associated with the dynamics of the universe. For the metric of the form eq. (\ref{eq4}) and considering a perfect fluid in a flat geometry, the Ricci scalar and Gauss-Bonnet parameter are presented as
\begin{eqnarray}
 && R=6\Big(2H^{2}+\dot{H}\Big),\\
 && G=24H^{4}\Big(1+\frac{\dot{H}}{H^{2}}\Big),\label{eq6}
\end{eqnarray}
where $H=\frac{\dot{a}}{a}$ is the Hubble parameter. The dot describes differentiation with respect to cosmic time $t$.  The variation principle in the gravitational action with respect to the metric $g^{ij}$, and $G$  produces the field equations for $f(G)$ gravity which can be presented as
\begin{eqnarray}
 && 3H^{2}=\rho_{m}+\rho_{G},\label{eq7}\\
 && -3H^{2}-2\dot{H}=p_{m}+p_{G}\label{eq8},
 \end{eqnarray}
where  $\rho_{m}$ and $p_{m}$ represent both relativistic matter (photons, neutrons) and non-relativistic matter (baryons, leptons, Cold Dark Matter ), $\rho_{G}$ and $p_{G}$  represent the contribution from the Gauss-Bonnet term.
The curvature fluid thermodynamical quantities of the Gauss-Bonnet gravity  for zero-order, namely curvature-energy density  and the isotropic pressure are presented as \cite{li2007cosmology,venikoudis2022late}
\begin{eqnarray}
&&\rho_{G}=\frac{Gf'-f}{2}-12\dot{f}'H^{3},\label{eq9}\\
&&p_{G}=\frac{f-Gf'}{2}+8\dot{f}'H\dot{H}+4H^{2}\left(\ddot{f}'+2H\dot{f}'\right)\label{eq10}.
\end{eqnarray}
The matter follows the perfect fluid assumption for any cosmological era of interest. The pressure of the perfect fluid is presented as
\begin{eqnarray}
 p_{m}=w\rho_{m},
\end{eqnarray}
 where $m$ specifies relativistic or non-relativistic matter, whereas the pressure of the total fluids is given by
\begin{eqnarray}
p_{t}=w_{t}\rho_{t}\;,
\end{eqnarray}
where $w_{t}=w_{m}+w_{G}=\frac{p_{m}+p_{G}}{\rho_{m}+\rho_{G}}$ and $\rho_{t}=\rho_{m}+\rho_{G}$.
The continuity equations in the context of normal matter and $f(G)$ gravity are given by
\begin{eqnarray}
&&\dot{\rho}_{m}+3H\Big(1+3w_{m}\Big)\rho_{m}=0\;,\label{eq13}\\
&&\dot{\rho}_{G}+3H\Big(1+3w_{G}\Big)\rho_{G}=0\;,\label{eq14}
\end{eqnarray}
with the equation of state parameter for matter fluid and Gauss-Bonnet fluid are given respectively as
\begin{eqnarray}
 w_{m}=\frac{p_{m}}{\rho_{m}}\;,
w_{G}=\frac{p_{G}}{\rho_{G}}\;.
\end{eqnarray}
Putting eq.(\ref{eq9}) and eq. (\ref{eq10}) into eq.(\ref{eq7}) and eq. (\ref{eq8})  produces
\begin{eqnarray}
 && 3H^{2}=\rho_{m}+\frac{Gf'-f}{2}-12f''\dot{G}H^{3},\label{eq16}\\
 && -3H^{2}-2\dot{H}=P_{m}+\frac{f-Gf'}{2}+8f''H^{3}\dot{G},\label{eq17}
 \end{eqnarray} which are the Friedmann and Raychaudhuri equations in the context of $f(G)$ gravity.
\subsection{Dynamical system of matter-Gauss-Bonnet fluids mixture}
The eq. (\ref{eq16}) can be rewritten as
\begin{equation}
1=\frac{\rho_{m}}{3H^{2}}+\frac{Gf'-f}{6H^{2}}-4f''\dot{G}H,\label{eq18}
\end{equation} which produces
\begin{eqnarray}
1+y-x-(\Omega_{d}+\Omega_{r})=0,\label{eq19}
\end{eqnarray} which is a dimensionless Friedmann equation,
where \begin{eqnarray}
&&\Omega_{d}=\frac{\rho_{d}}{3H^{2}},\label{eq20}\\
&& \Omega_{r}=\frac{\rho_{r}}{3H^{2}},\label{eq21}\\
&& x=\frac{Gf'-f}{6H^{2}},\label{eq22}\\
&& y=4f''\dot{G}H.\label{eq23}
\end{eqnarray} In this process, we have considered dust energy density $\rho_{d}$ and radiation energy density $\rho_{r}$ separately in a flat Universe.  One of the simplest form of $f(G)$ models is the power-law model given by \cite{bamba2010finite}
\begin{eqnarray}
&&f=\beta H^{4}_{0}\Big(\frac{G}{H^{4}_{0}}\Big)^{n},\label{eq24}
\end{eqnarray} with this model, we get
\begin{eqnarray}
&& x=\beta(n-1)H^{4-4n}_{0}\frac{G^{n}}{6H^{2}},\label{eq25}\\
&& y=4\beta H^{4-4n}_{0}n(n-1)G^{n-2}\dot{G}H.\label{eq26}
\end{eqnarray}
 The eq. (\ref{eq17}) can then be represented as (after using eq. (\ref{eq20}), eq. (\ref{eq21}), eq. (\ref{eq25}) and eq. (\ref{eq26}))
\begin{eqnarray}
&& \frac{\dot{H}}{H^{2}}=-\frac{3}{2}-\frac{3}{2}(\Omega_{r}+\Omega_{d})+\frac{3}{2}x-y.\label{eq27}
\end{eqnarray}
Defining $N=\ln (a)$ \cite{carloni2005cosmological}, where $\frac{df}{Hdt}=\frac{df}{dN}$, the evolution of dimensionless variables $x$, $y$, $\Omega_{d}$ and $\Omega_{r}$ (eq. (\ref{eq20}), eq. (\ref{eq21}), eq. (\ref{eq25}) and eq. (\ref{eq26})) can be presented as 
\begin{eqnarray}
&&\frac{d \Omega_{r}}{dN}=(3w)\Omega_{r}+\Big((2+3w)y-3(w+1)x\Big)\Omega_{r},\label{eq28}\\
&& \frac{d \Omega_{d}}{dN}=(3w)\Omega_{d}+(2+3w)y\Omega_{d}-3x(w+1)\Omega_{d},\label{eq29}\\
&& \frac{d {h}}{dN}=h\Big[-\frac{3}{2}-\frac{3w}{2}(1+y-x)+\frac{3}{2}x-y\Big],\label{eq30}\\
&& \frac{d {x}}{dN}=3(1+w)x-3(1+w)x^{2}+\Big[\frac{9w+5}{2}-\frac{1+3w}{2}-(1+\frac{3w}{2})y\big]y,\label{eq31}\\
&& \frac{d {y}}{dN}=\Big[\frac{3x}{2}x-\frac{3}{2}-\frac{3w(1+y-x)}{2}\Big]y-\Big[1+\frac{(n-2)}{nx}\Big(\frac{1}{2}+\frac{3w}{2}(1+y-x)\nonumber\\&&-\frac{3x}{2}-y\Big)\Big]y^{2},\label{eq32}
\end{eqnarray}
where eq. (\ref{eq30}) was obtained from $h=\frac{H}{H_{0}}$. The stability of the equilibrium points of dynamical system in $f(G)$ has been studied in details in \cite{khyllep2207cosmology} and in \cite{singh2025dynamical} for $f(G)$ gravity, whereas the matter power spectrum has been studied in \cite{abebe2013large,ntahompagaze2022large,fedeli2012matter}. In this study, we study the matter power spectrum in a power-law $f(G)$ gravity. To this end, let us first present eqs. (\ref{eq28})--eq. (\ref{eq32}) into redshift space. Starting from $a=\frac{1}{1+z}$, then for any quantity $\mathcal{K}(z)$, one can write $\frac{d\mathcal{K}}{dN}=-(1+z)\frac{d\mathcal{K}}{dz}$. Therefore eqs. (\ref{eq28})--eq. (\ref{eq32}) can be represented in redshift space as
\begin{eqnarray}
&&-(1+z)\frac{d \Omega_{r}}{dz}=(3w)\Omega_{r}+\Big((2+3w)y-3(w+1)x\Big)\Omega_{r},\label{eq33}\\
&& -(1+z)\frac{d \Omega_{d}}{dz}=(3w)\Omega_{d}+(2+3w)y\Omega_{d}-3x(w+1)\Omega_{d},\label{eq34}\\
&& -(1+z)\frac{d {h}}{dz}=h\Big[-\frac{3}{2}-\frac{3w}{2}(1+y-x)+\frac{3}{2}x-y\Big],\label{eq35}\\
&& -(1+z)\frac{d {x}}{dz}=3(1+w)x-3(1+w)x^{2}+\Big[\frac{9w+5}{2}-\frac{1+3w}{2}-(1+\frac{3w}{2})y\big]y,\label{eq36}\\
&& -(1+z)\frac{d {y}}{dz}=\Big[\frac{3x}{2}x-\frac{3}{2}-\frac{3w(1+y-x)}{2}\Big]y-\Big[1+\frac{(n-2)}{nx}\Big(\frac{1}{2}+\frac{3w}{2}(1+y-x)\nonumber\\&&-\frac{3x}{2}-y\Big)\Big]y^{2}.\label{eq37}
\end{eqnarray} Eqs. (\ref{eq33})--(\ref{eq37}) will be used together with perturbation equations to compute matter-Gauss-Bonnet power spectrum.
\section{Covariant density perturbations in $f(G)$ gravity}\label{sec3}
In this paper, we consider the $1+3$ covariant formalism, where the space-time is split into temporal and spatial components with respect to the congruence. This formalism treats the slicing of four dimensional space by time and hyper-surface. For congruence normal to a space-like hyper-surface, the $1+3$ covariant formalism reduces to $3+1$ \cite{gourgoulhon20073+}. The basic structure of the $1+3$ covariant formalism is a congruence of one dimensional curves, mostly time-like curves \cite{park2018covariant,gourgoulhon20073+}. In this case, we decompose space-time cosmological manifold into time and space sub-manifold separately with a perpendicular $4$-velocity field vector $u^{a}$ so that
\begin{eqnarray}
 u^{a}=\frac{dx^{a}}{d\tau}\;,
\end{eqnarray}
with the metric tensor related to the spatial component as
\begin{eqnarray}
 g_{\mu\nu}=h_{ab}-u_{a}u_{b}.
\end{eqnarray}
The scalar part of perturbations play a key role in matter clustering hence large scale structure formation. Thus we extract the scalar part from the vector gradient quantities using the local decomposition for a vector quantity $X_{a}$ as \cite{abebe2012covariant}
\begin{eqnarray}
a\tilde{\bigtriangledown}_{b}X_{a}=X_{ab}=\frac{1}{3}h_{ab}X+\sum_{ab}X+X_{[ab]},
\end{eqnarray} where $\sum_{ab}X=X_{(ab)}-\frac{1}{3}h_{ab}X$ describes the shear and $X_{[ab]}$ describes the vorticity.
  We define the gradient variables to obtain perturbation  equations governing the evolution of the universe as 
\begin{eqnarray}
\Delta_{m}=a^{2}\frac{\tilde{\bigtriangledown}^{2}\rho_{m}}{\rho_{m}}, Z=a^{2}\tilde{\bigtriangledown}^{2}\theta,
\mathcal{G}=a^{2}\tilde{\bigtriangledown}^{2}G, and 
 \mathcal{C}=a^{2}\tilde{\bigtriangledown}^{2}\dot{G}
\end{eqnarray} and by using scalar and harmonic decomposition techniques presented as
\cite{abebe2012covariant,munyeshyaka2021cosmological}:
For a given quantity $X$, one writes
 \begin{eqnarray}
  X=\sum_{k}X^{k}(t).Q_{k}(\vec{x}),
 \end{eqnarray}
where
\begin{eqnarray}
 &&\tilde{\bigtriangledown}^{2}=-\frac{k^{2}}{a^{2}}Q^{k},~ \dot{Q}_{k}(\vec{x})=0,~ k=\frac{2 \pi a}{\lambda},
\end{eqnarray}
with $Q^{k}$, $k$ and $\lambda$ are the eigenfunction of the co-moving spatial Laplace-Beltrami operator, the order of harmonic (wave-number) and the physical wavelength of the mode, respectively.
We present  linear evolution equations of the matter energy-density perturbation as 
\begin{eqnarray}
&&\ddot{\Delta}^{k}_{m}=\Big[w\theta -\frac{2}{3}\theta-6f'''\dot{G}H^{2}\Big]\dot{\Delta}^{k}_{m}+\Big[ w\Big(-\frac{1}{2}(1+3w_{m})\rho_{m}+6f''\dot{G}H^{3}+6f''GH^{3}\nonumber\\&&+12f'''\dot{G}H^{3}\Big)+\frac{(1+w)}{2}(1+3w)\rho_{m}+w\frac{k^{2}}{a^{2}}\Big]\Delta^{k}_{m}+(1+w)6H^{3}f''\dot{\mathcal{G}}^{k}\nonumber\\&&-(1+w)\Big[\frac{f'+f''-1}{2}-6\dot{G}f'''H^{3}+\frac{k^{2}}{a^{2}}\Big]\mathcal{G}^{k},\label{eq44}\\
&&\ddot{\mathcal{G}}^{k}=-\frac{2w\dot{G}}{1+w}\Delta^{k}_{m}-\frac{wG}{1+w}\dot{\Delta}^{k}_{m}.\label{eq45}
\end{eqnarray}

Throughout this work, we assume $w_{m}$ to be a constant  and $w_{G}$  dynamically changes. The Raychaudhuri equation $\dot{\theta}=-\frac{1}{3}\theta^{2}-\frac{1}{2}(\rho+3p)+\bigtriangledown_{a}\dot{u}_{a}$  that governs the expansion history of the universe in this case is given by
$\dot{\theta}=-\frac{1}{3}\theta^{2}-\frac{1}{2}\Big((1+3w_{m})\rho_{m}+Gf'-f-12f''\dot{G}H^{3}\Big)+\bigtriangledown_{a}\dot{u}_{a}$,
with $\dot{u}_{a}$ as the $4$-acceleration in the energy frame of the total fluids  given by
$\dot{u}_{a}=-\frac{1}{(1+w)\rho_{m}}\Big[w_{m}\tilde{\bigtriangledown}_{a}\rho_{m}\Big]$. Most of large scale structure are believed to have formed  during matter domination epoch, after decoupling, thus we consider the equation of state parameter to be that of dust ($w=0$) so that equations (\ref{eq44})--(\ref{eq45}) are rewritten as 
 for dust $w=0$
\begin{eqnarray}
&&\ddot{\Delta}^{k}_{m}=\Big[-\frac{2}{3}\theta-6f'''\dot{G}H^{2}\Big]\dot{\Delta}^{k}_{m}+\Big[ \frac{1}{2}\rho_{d}\Big]\Delta^{k}_{m}+6H^{3}f''\dot{\mathcal{G}}^{k}\nonumber\\&&-\Big[\frac{f'+f''-1}{2}-6\dot{G}f'''H^{3}+\frac{k^{2}}{a^{2}}\Big]\mathcal{G}^{k},\label{eq46}\\
&&\ddot{\mathcal{G}}^{k}=-\frac{2w_{d}\dot{G}}{1+w}\Delta^{k}_{m}-\frac{w_{d}G}{1+w}\dot{\Delta}^{k}_{m}.\label{eq47}
\end{eqnarray} To analyse the obtained perturbation equations for the dust dominated universe, we transfrom them into dimensionless form  and present them into redshift space using
$ a=\frac{a_{0}}{1+z}$, $\dot{f}=-(1+z)Hf'$, $\ddot{f}=(1+z)^{2}H\Big[H'f'+Hf''\Big]+(1+z)H^{2}f'$,
        where $'$ means differentiation with respect to redshift. Throughout this manuscript, $a_{0}$ is set to $1$. The perturbation equations eqs (\ref{eq46})--(\ref{eq47}) are therefore represented in redshift space as
\begin{eqnarray}
&&\Delta''_{m}=-\frac{1}{(1+z)}\Big[-2-\frac{3(n-2)}{2q}y\Big(1+z\Big)^{-\frac{6}{m}}+\frac{3}{2m}+1\Big]\Delta'_{m}+\frac{3\Omega_{d}}{(1+z)^{2}}\Delta_{m}\nonumber\\&&-\frac{\Big(-\frac{3}{2}-\frac{3}{2}\Omega_{d}+\frac{3}{2}x-y\Big)}{2q}3nx(1+z)^{-\frac{6}{m}-1}\mathcal{G}'\nonumber\\&&-\frac{1}{(1+z)^{2}}\Big[\frac{3x\Big(1+z\Big)^{-\frac{6}{m}}}{(n-1)q}\Big[1+\frac{(n-1)}{q}\Big(1+z\Big)^{-\frac{6}{m}}\Big]-\frac{3}{4m}\Big(1+z\Big)^{-\frac{3}{2m}}\nonumber\\&&-\frac{3(n-2)y}{2q}\Big(1+z\Big)^{-\frac{6}{m}}\Big]+\frac{k^{2}}{a^{2}}\Big]\mathcal{G}\label{eq48}\\
&&\mathcal{G}''=-\frac{1}{(1+z)}\Big[\frac{3}{2m}+1\Big]\mathcal{G}',\label{eq49}
\end{eqnarray}
where we have used eqs. (\ref{eq20})--(\ref{eq21}), eqs. (\ref{eq25})--(\ref{eq26}).
\section{Large scale structure power spectrum in the context of $f(G)$ gravity}\label{sec4}
Having obtained the perturbation equations in the dust dominated universe, next step is to compute the power spectrum from the obtained equations. Assuming an isotropic FRW universe, it is shown that $P^{f(G)}_{k}$ which can be compared to the $\Lambda$CDM predictions is given by \cite{abebe2013large}
\begin{eqnarray}
P^{f(G)}_{k}=\left|\Delta(k)\right|^{2}.
\end{eqnarray}
In order to compute the power specta, we considered $5$ different initial conditions for the system of equations (\ref{eq48})--(\ref{eq49}), in order to study the sensitivity on the processed spectrum.
\begin{itemize}
\item Set I: $\Delta_{d}(zin=2000)=10^{-2}$, $\Delta'_{d}(zin=2000)=10^{-8}$, $\mathcal{G}(zin=2000)=10^{-2}$ and $\mathcal{G}'(zin=2000)=10^{-8}$,
\item  Set II: $\Delta_{d}(zin=2000)=10^{-5}$, $\Delta'_{d}(zin=2000)=10^{-5}$, $\mathcal{G}(zin=2000)=10^{-5}$ and $\mathcal{G}'(zin=2000)=10^{-5}$,
\item Set III: $\Delta_{d}(zin=2000)=10^{-5}$, $\Delta'_{d}(zin=2000)=0$, $\mathcal{G}(zin=2000)=10^{-5}$ and $\mathcal{G}'(zin=2000)=0$, 
\item set IV: $\Delta_{d}(zin=2000)=10^{-5}$, $\Delta'_{d}(zin=2000)=10^{-8}$, $\mathcal{G}(zin=2000)=10^{-5}$ and $\mathcal{G}'(zin=2000)=10^{-8}$ and 
\item set V: $\Delta_{d}(zin=2000)=10^{-5}$, $\Delta'_{d}(zin=2000)=10^{-3}$, $\mathcal{G}(zin=2000)=10^{-5}$ and $\mathcal{G}'(zin=2000)=10^{-3}$,
\end{itemize} 
where $zin$ is the initial redshift during radiation and matter equality. The choice of the these sets of initial conditions for the system of equations can be understood as providing scale invariant for $\Delta_{m}$ and $\mathcal{G}$ and there first derivatives \cite{abebe2013large}. To obtain the results, we have solved simultaneously eq. (\ref{eq34}), eq. (\ref{eq36}) and eq. (\ref{eq37}) to get the redshift dependent solutions for parameter $x$, $y$ and $\Omega_{d}$. We then used to obtained solutions to solve the system of eq. (\ref{eq48})--(\ref{eq49}), to obtain the energy density perturbations $\Delta_{d}(z)$ which then used to compute the power spectrum. The integration was done at a fixed redshift, say today ($z=0$) and the power spectrum $P^{f(G)}_{k}$ was computed against $k$ and presented in the following plots: 
\begin{figure}
 \includegraphics[width=85mm,height=60mm]{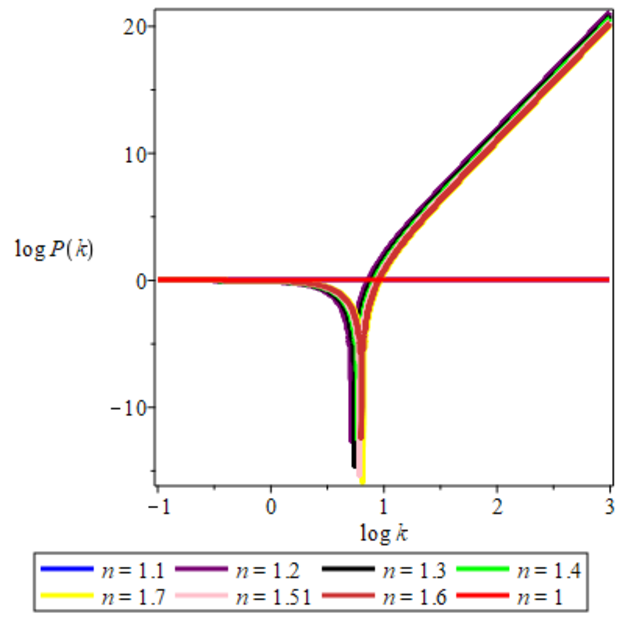}
\caption{Matter power spectrum of eqs. (\ref{eq48})--(\ref{eq49}) for different values of $n$ using the  initial conditions $\Delta_{d}(zin=2000)=10^{-2}$, $\Delta'_{d}(zin=2000)=10^{-8}$, $\mathcal{G}(zin=2000)=10^{-2}$ and $\mathcal{G}'(zin=2000)=10^{-8}$ were used. For $n=1$, GR case is recoverd.}
  \label{Fig1}
 \end{figure}
\begin{figure}
 \includegraphics[width=85mm,height=60mm]{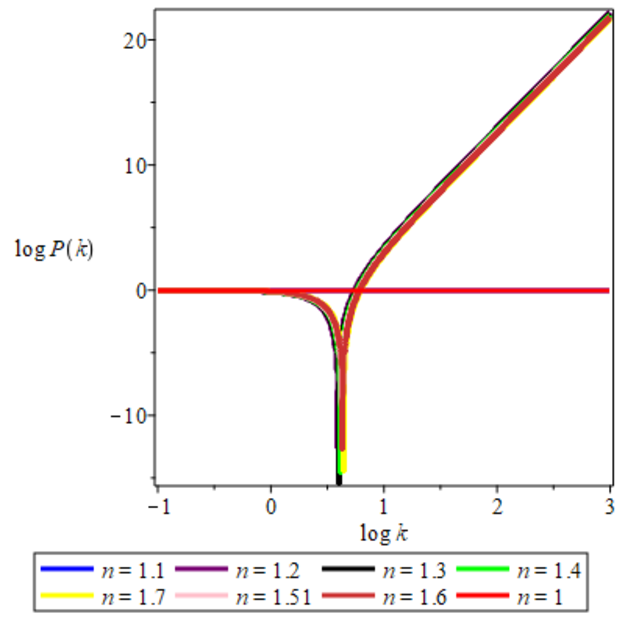}
\caption{Matter power spectrum of eqs. (\ref{eq48})--(\ref{eq49}) for different values of $n$ using  the  initial conditions Set II: $\Delta_{d}(zin=2000)=10^{-5}$, $\Delta'_{d}(zin=2000)=10^{-5}$, $\mathcal{G}(zin=2000)=10^{-5}$ and $\mathcal{G}'(zin=2000)=10^{-5}$  were used. For $n=1$, GR case is recoverd.}
  \label{Fig2}
	\end{figure}
	\begin{figure}
 \includegraphics[width=85mm,height=60mm]{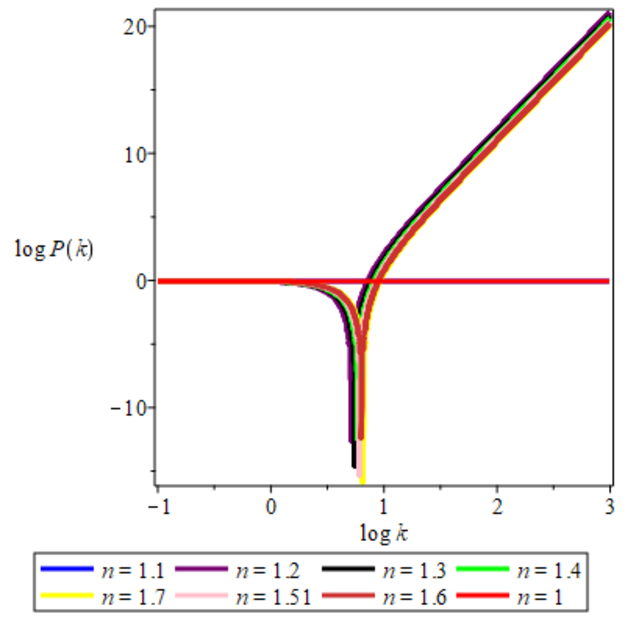}
\caption{Matter power spectrum of eqs. (\ref{eq48})--(\ref{eq49}) for different values of $n$ using  the  initial conditions $\Delta_{d}(zin=2000)=10^{-5}$, $\Delta'_{d}(zin=2000)=0$, $\mathcal{G}(zin=2000)=10^{-5}$ and $\mathcal{G}'(zin=2000)=0$ were used. For $n=1$, GR case is recoverd.}
  \label{Fig3}
 \end{figure}
\begin{figure}
 \includegraphics[width=85mm,height=60mm]{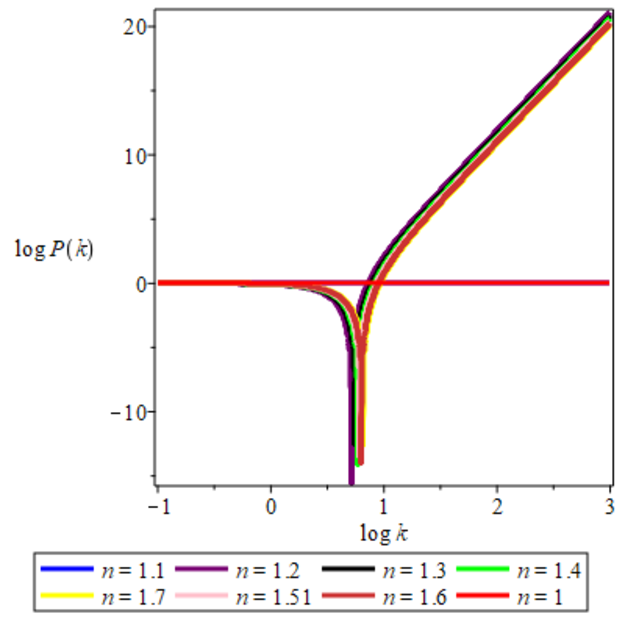}
\caption{Matter power spectrum of eqs. (\ref{eq48})--(\ref{eq49}) for different values of $n$.  To find numerical results, the  initial conditions  $\Delta_{d}(zin=2000)=10^{-5}$, $\Delta'_{d}(zin=2000)=10^{-8}$, $\mathcal{G}(zin=2000)=10^{-5}$ and $\mathcal{G}'(zin=2000)=10^{-8}$  were used. For $n=1$, GR case is recoverd.}
  \label{Fig4}
	\end{figure}
	\begin{figure}
 \includegraphics[width=85mm,height=60mm]{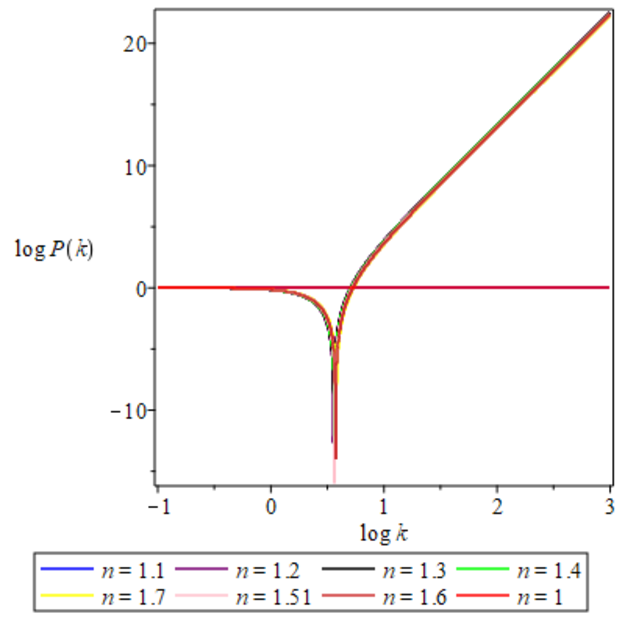}
\caption{Matter power spectrum of eqs. (\ref{eq48})--(\ref{eq49}) for different values of $n$.  To find numerical results, the  initial conditions $\Delta_{d}(zin=2000)=10^{-5}$, $\Delta'_{d}(zin=2000)=10^{-3}$, $\mathcal{G}(zin=2000)=10^{-5}$ and $\mathcal{G}'(zin=2000)=10^{-3}$ were used. For $n=1$, GR case is recoverd.}
  \label{Fig5}
 \end{figure}
 \begin{figure}
 \includegraphics[width=85mm,height=60mm]{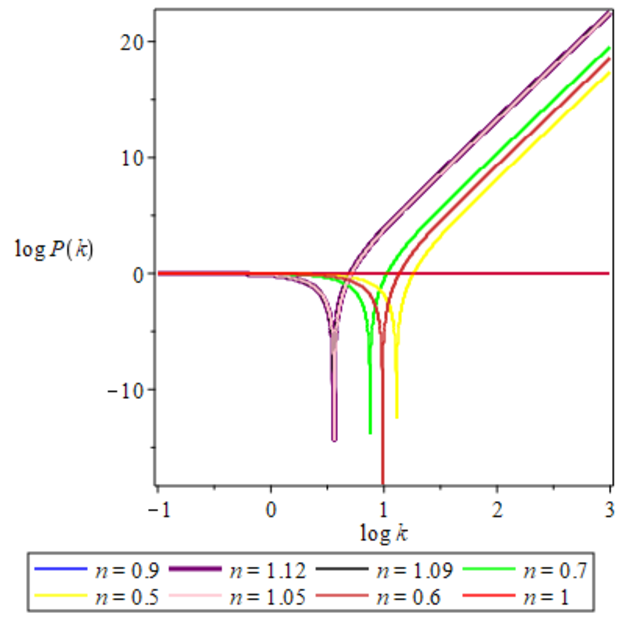}
\caption{Matter power spectrum for different values of $n$ using  eqs. (\ref{eq48})--(\ref{eq49}). To find numerical results, the  initial conditions $\Delta(z_{0})=10^{-5}$, $\Delta'(z_{0})=0$, $\mathcal{G}(z_{0})=10^{-5}$ and $\mathcal{G}'(z_{0})=0$ were used. For $n=1$, GR case is recoverd. }
  \label{Fig6}
 \end{figure}
\clearpage
\section{Discussions and conclusions}\label{sec5}
\subsection{Discussions}
In the present paper, we presented the energy density perturbations and computed the corresponding matter power spectrum for $f(G)$ gravity model.  For computing the matter power spectrum, we consider different values of the power-law index $n$ in the considered $f(G)=\beta G^{n}$ model. The figures were obtained after considering five different set of initial conditions following the work presented in \cite{ntahompagaze2022large,abebe2013large} at the initial redshift $z0=2000$, namely: Set I: $\Delta_{d}(zin=2000)=10^{-2}$, $\Delta'_{d}(zin=2000)=10^{-8}$, $\mathcal{G}(zin=2000)=10^{-2}$ and $\mathcal{G}'(zin=2000)=10^{-8}$, Set II: $\Delta_{d}(zin=2000)=10^{-5}$, $\Delta'_{d}(zin=2000)=10^{-5}$, $\mathcal{G}(zin=2000)=10^{-5}$ and $\mathcal{G}'(zin=2000)=10^{-5}$, Set III: $\Delta_{d}(zin=2000)=10^{-5}$, $\Delta'_{d}(zin=2000)=0$, $\mathcal{G}(zin=2000)=10^{-5}$ and $\mathcal{G}'(zin=2000)=0$, set IV: $\Delta_{d}(zin=2000)=10^{-5}$, $\Delta'_{d}(zin=2000)=10^{-8}$, $\mathcal{G}(zin=2000)=10^{-5}$ and $\mathcal{G}'(zin=2000)=10^{-8}$ and set V: $\Delta_{d}(zin=2000)=10^{-5}$, $\Delta'_{d}(zin=2000)=10^{-3}$, $\mathcal{G}(zin=2000)=10^{-5}$ and $\mathcal{G}'(zin=2000)=10^{-3}$.  The plots for matter power spectrum are presented in fig. (\ref{Fig1}) for Set I, fig. (\ref{Fig2}) for set II, fig. (\ref{Fig3}) for set III, fig. (\ref{Fig4}) for set IV and fig. (\ref{Fig5}) for set IV. The effects of setting these sets of initial conditions were stressed in the work done by \cite{abebe2013large}.  The energy density perturbations were evaluated at the present day redshift $z=0$ and we set  ranges of the wave-number $k$ basing on the work done in \cite{tegmark2006cosmological}. For all the sets of initial conditions, the matter power spectrum does not presents an oscillation behavior and the curves almost behave similary for  different values of $n$ considered. A result differs to the one Obtained in \cite{ntahompagaze2022large}. For $n=1$, the GR case is recovered, where the matter power spectrum does not depends on the value of $k$ .  As $n$ gets smaller as presented in fig. (\ref{Fig6}), the curves separate and the power spectrum decays and evolves above the GR invariant line. In the present work, one can highlight that i) The obtained matter energy density ($\Delta$) in a dust dominated universe couples with the energy density resulting from Gauss-Bonnet contribution ($\mathcal{G}$), therefore $\mathcal{G}$ influences the behavior of the power spectrum. ii) The obtained power spectra decays with a change in amplitude then evolve above the invariant line (GR) as $k$ increases for different values of $n$ but as $n$ gets smaller (closer to $1$), the curves start to separates from each other, as result not identified in \cite{ntahompagaze2022large}. iii) For all the set of initial conditions, the behavior of the power spectrum does not show oscillations. This characteristic is in agreement with the work conducted in \cite{abebe2013large,fedeli2012matter}
\subsection{Conclusion}
In the present paper, an analysis of the background and matter perturbations for $f(G)=G^{n}$ gravity model is presented. After defining gradient variables and using both dynamical system approach and the $1+3$ covariant formalism, we derived and solved the linear perturbation equations responsible for large scale structure formation. This was done to study the evolution of matter density perturbations  in a dust dominated universe without considering the quasi-static approximation. We used different values of exponent $n$ and different sets of initial conditions to obtain the energy density perturbations which then used to compute power spectra for a particular functional form of $f(G)$ model in the context of $f(G)$ gravity. By looking at the plots, the obtained spectra the curves decay and evolve above the invariant GR line as $k$ increases. In this case, the exponent giving a slight different results in terms of the behavior of the curves appear for $n$ closer to $1$. We can conclude that the results presented in this work in the context of $f(G)$ gravity support the $\Lambda$ CDM  and GR predictions. The future work should consider using different $f(G)$ models to constrain the results with observations. This will be done elsewhere. 
  
\section*{acknowledgements} AM acknowledges the hospitality of the Department of Physics of the University of Rwanda, where this work was conceptualized and completed.  PKD gratefully acknowledge the support from IUCAA, Pune, India provided through the Visiting Research Associateship Program. JN thanks the support from Rwanda Astrophysics, Space and Climate Science research Group.

\noindent
{\color{blue} \rule{\linewidth}{1mm} }

\begin{thebibliography}{00}


\bibitem{perlmutter1999astrophys}
Perlmutter S, Collaboration S~C~P {\em et~al.\/} 1999 {\em Astron. J\/} {\bf
  116} 1009

\bibitem{dodelson2000dark}
Dodelson S and Knox L 2000 {\em Physical Review Letters\/} {\bf 84} 3523

\bibitem{cornish2004constraining}
Cornish N~J, Spergel D~N, Starkman G~D and Komatsu E 2004 {\em Physical Review
  Letters\/} {\bf 92} 201302

\bibitem{spergel2007three}
Spergel D~N, Bean R, Dor{\'e} O, Nolta M, Bennett C, Dunkley J, Hinshaw G,
  Jarosik N~e, Komatsu E, Page L {\em et~al.\/} 2007 {\em The astrophysical
  journal supplement series\/} {\bf 170} 377

\bibitem{tegmark2004cosmological}
Tegmark M, Strauss M~A, Blanton M~R, Abazajian K, Dodelson S, Sandvik H, Wang
  X, Weinberg D~H, Zehavi I, Bahcall N~A {\em et~al.\/} 2004 {\em Physical
  review D\/} {\bf 69} 103501

\bibitem{singh2020cosmological}
Singh G, Hulke N and Singh A 2020 {\em Indian Journal of Physics\/} {\bf 94}
  127--141

\bibitem{de2010theories}
De~Felice A and Tsujikawa S 2010 {\em Living Reviews in Relativity [electronic
  only]\/} {\bf 13} Article--No

\bibitem{sotiriou2010f}
Sotiriou T~P and Faraoni V 2010 {\em Reviews of Modern Physics\/} {\bf 82}
  451--497

\bibitem{gomez2011standard}
Gomez F, Minning P and Salgado P 2011 {\em Physical Review D—Particles,
  Fields, Gravitation, and Cosmology\/} {\bf 84} 063506

\bibitem{bajardi2021exact}
Bajardi F, Vernieri D and Capozziello S 2021 {\em Journal of Cosmology and
  Astroparticle Physics\/} {\bf 2021} 057

\bibitem{tseytlin1997non}
Tseytlin A~A 1997 {\em Nuclear Physics B\/} {\bf 501} 41--52

\bibitem{garcia2011energy}
Garcia N~M, Harko T, Lobo F~S and Mimoso J~P 2011 {\em Physical Review
  D—Particles, Fields, Gravitation, and Cosmology\/} {\bf 83} 104032

\bibitem{li2007cosmology}
Li B, Barrow J~D and Mota D~F 2007 {\em Physical Review D—Particles, Fields,
  Gravitation, and Cosmology\/} {\bf 76} 044027

\bibitem{de2020tracing}
de~Martino I, De~Laurentis M and Capozziello S 2020 {\em Physical Review D\/}
  {\bf 102} 063508

\bibitem{benetti2018observational}
Benetti M, Santos~da Costa S, Capozziello S, Alcaniz J~S and De~Laurentis M
  2018 {\em International Journal of Modern Physics D\/} {\bf 27} 1850084

\bibitem{de2012stability}
De~la Cruz-Dombriz A and S{\'a}ez-G{\'o}mez D 2012 {\em Classical and Quantum
  Gravity\/} {\bf 29} 245014

\bibitem{munyeshyaka2024covariant}
Munyeshyaka A, Ntahompagaze J, Mutabazi T and Mbonye M 2024 {\em The European
  Physical Journal C\/} {\bf 84} 51

\bibitem{bamba2012cosmic}
Bamba K, Lopez-Revelles A, Myrzakulov R, Odintsov S and Sebastiani L 2012 {\em
  Classical and Quantum Gravity\/} {\bf 30} 015008

\bibitem{bamba2011phantom}
Bamba K, Geng C~Q and Lee C~C 2011 {\em International Journal of Modern Physics
  D\/} {\bf 20} 1339--1345

\bibitem{wainwright1997dynamical}
Wainwright J and Ellis G~F~R 1997 {\em Dynamical systems in cosmology\/}

\bibitem{bohmer2017dynamical}
B{\"o}hmer C~G and Chan N 2017 Dynamical systems in cosmology {\em Dynamical
  and Complex Systems\/} (World Scientific) pp 121--156

\bibitem{singh2025dynamical}
Singh A 2025 {\em The European Physical Journal C\/} {\bf 85} 1--12

\bibitem{ntahompagaze2022large}
Ntahompagaze J, Abebe A and Mbonye M~R 2022 {\em International Journal of
  Modern Physics D\/} {\bf 31} 2250071

\bibitem{abebe2013large}
Abebe A, de~la Cruz-Dombriz A and Dunsby P~K 2013 {\em Physical Review
  D—Particles, Fields, Gravitation, and Cosmology\/} {\bf 88} 044050

\bibitem{venikoudis2022late}
Venikoudis S, Fasoulakos K and Fronimos F 2022 {\em International Journal of
  Modern Physics D\/} {\bf 31} 2250038

\bibitem{bamba2010finite}
Bamba K, Odintsov S~D, Sebastiani L and Zerbini S 2010 {\em The European
  Physical Journal C\/} {\bf 67} 295--310

\bibitem{carloni2005cosmological}
Carloni S, Dunsby P~K, Capozziello S and Troisi A 2005 {\em Classical and
  Quantum Gravity\/} {\bf 22} 4839

\bibitem{khyllep2207cosmology}
Khyllep W {\em arXiv preprint arXiv:2207.02610\/}

\bibitem{fedeli2012matter}
Fedeli C, Dolag K and Moscardini L 2012 {\em Monthly Notices of the Royal
  Astronomical Society\/} {\bf 419} 1588--1602

\bibitem{gourgoulhon20073+}
Gourgoulhon E 2007 {\em arXiv preprint gr-qc/0703035\/}

\bibitem{park2018covariant}
Park C 2018 {\em arXiv preprint arXiv:1810.06293\/}

\bibitem{abebe2012covariant}
Abebe A, Abdelwahab M, De~la Cruz-Dombriz A and Dunsby P~K 2012 {\em Classical
  and quantum gravity\/} {\bf 29} 135011

\bibitem{munyeshyaka2021cosmological}
Munyeshyaka A, Ntahompagaze J and Mutabazi T 2021 {\em International Journal of
  Modern Physics D\/} {\bf 30} 2150053

\bibitem{tegmark2006cosmological}
Tegmark M, Eisenstein D~J, Strauss M~A, Weinberg D~H, Blanton M~R, Frieman J~A,
  Fukugita M, Gunn J~E, Hamilton A~J, Knapp G~R {\em et~al.\/} 2006 {\em
  Physical Review D—Particles, Fields, Gravitation, and Cosmology\/} {\bf 74}
  123507
\end{thebibliography}
\end{document}